\documentclass[preprint,aps]{revtex4}
\usepackage{dcolumn,graphicx}
\def\be {\begin{equation}}
\def\ee {\end{equation}}
\def\mn {{\mu\nu}}
\def\ba {\begin{eqnarray}}
\def\ea {\end{eqnarray}}
\def\nn {\nonumber}
\def\cm {{\cal M}}
\def\cl {{\cal L}}
\def\la {\langle}
\def\ra {\rangle}
\def\del {\partial}

\def\om {\omega}
\def\gm {\gamma}

\def\vq {\vec q}
\def\vk {\vec k}
\def\vp {\vec p}
\def\ep {\epsilon}
\def\de {\delta}
\def\Gm {\Gamma}
\def\De {\Delta}
\def\sg {\sigma}
\def\omp {\om_\pi}
\def\omh {\om_h}

\begin{document}
\title{Medium effects on the viscosities of a pion gas}
\author{Sukanya Mitra}
\author{Sourav Sarkar}
\affiliation{Theoretical Physics Division, Variable Energy Cyclotron Centre, 1/AF Bidhannagar
Kolkata - 700064, India}
\begin{abstract}
The bulk and shear viscosities of a pion gas is obtained by solving the
relativistic transport equation in the Chapman-Enskog approximation. In-medium
effects are introduced in the $\pi\pi$ cross-section through one-loop
self-energies in the propagator of the exchanged $\rho$ and $\sigma$ mesons. 
The effect of early
chemical freeze-out in heavy ion collisions is implemented through a temperature
dependent pion chemical potential. These are found to affect the temperature
dependence of the bulk and shear viscosities significantly.
\end{abstract}

\maketitle

\section{Introduction}

The study of transport coefficients and non-equilibrium dynamics in general has assumed a
great significance in recent times. The impetus in this direction has been
provided by the recent results from the Relativistic Heavy Ion Collider (RHIC),
in particular the elliptic flow which seems to be rather well described by nearly
ideal hydrodynamics with a very low value of $\eta/s$, close to the quantum
bound $1/4\pi$~\cite{KSS}, $\eta$ and $s$ being the coefficient of shear viscosity and
entropy density respectively. This led to the description of quark gluon plasma
as the most perfect fluid known~\cite{Csernai}. This bound is however found to
be violated, e.g. in the case of a strongly coupled anisotropic plasma~\cite{Rebhan}.

Of the two viscosities used to parametrize the leading order corrections to the
stress tensor the shear viscosity $\eta$ is more commonly discussed, the
magnitude of $\zeta$ the bulk viscosity usually being much smaller in
comparison. The latter actually vanishes as in the case of systems with
conformal invariance e.g. a gas of non-interacting highly relativistic
particles. Breaking of conformal invariance of QCD due to quantum
effects shows up through non-zero values of $\zeta$ which is related to the
trace of the energy momentum tensor. An estimate of this can be obtained
from the interaction measure determined from lattice calculations~\cite{Cheng}. 
The behaviour of $\zeta$ and $\eta$ as a function of temperature  
is particularly relevant in the context of non-ideal hydrodynamic
simulations of heavy ion collisions. Whereas $\eta/s$ as a function of $T$
is expected to go through a minimum~\cite{Csernai} at or near the critical value for crossover
from the partonic phase, it is believed that $\zeta/s$ may be large~\cite{Kharzeev}
or even diverging close to this value. These have been studied in the 
linear sigma model in the large-$N$ limit~\cite{Dobado1,Dobado2} and for a
massless pion gas in~\cite{Chen2}. 

The viscosities of a pion gas have received some attention in recent times.  
In the diagrammatic approach, one uses the Kubo formula which relates the
transport coefficients to retarded two-point functions~\cite{Zubarev,Sakagami}.
The shear viscosity of a pion gas in this approach was evaluated
in~\cite{Lang,Mallik_kubo}.
However, the kinetic theory approach being computationally more
efficient~\cite{Jeon},
has been mostly used to obtain the viscous coefficients. 
The $\pi\pi$ cross-section is a crucial
dynamical input in these calculations. Scattering amplitudes evaluated using chiral perturbation
theory~\cite{Weinberg1,Gasser} to lowest order have been used 
in~\cite{Santalla,Chen}
and unitarization improved estimates of the amplitudes were used 
in~\cite{Dobado3} to evaluate the shear viscosity. Phenomenological
scattering cross-section using experimental phase shifts have been
used in~\cite{Prakash,Chen,Davesne,Itakura}. 
While in~\cite{Moore} the effect of number changing processes
on the bulk viscosity of a pion gas has been studied, 
in~\cite{Fraile} unitarized chiral perturbation theory was
used to demonstrate the breaking of conformal symmetry by the pion
mass. 

Our aim in this work is to construct the $\pi\pi$ cross-section
in the medium and study its effect on the
temperature dependence of the viscous coefficients. 
For this purpose we employ an effective Lagrangian
approach which incorporates
$\rho$ and $\sigma$ meson exchange in $\pi\pi$ scattering.
A motivating factor is the role of the $\rho$ pole in $\pi\pi$ scattering 
in preserving the quantum bound on $\eta/s$ for a pion gas 
as demonstrated through use of the phenomenological~\cite{Bertsch1,Prakash} and
unitarized~\cite{Dobado3} cross-section
in~\cite{Dobado_kss}.  Medium effects are
then introduced in the $\rho$ and $\sigma$ propagators through one-loop self-energy diagrams.
Now, the hadronic matter produced in highly relativistic heavy ion collisions is
known to undergo early chemical freeze-out. Number changing (inelastic)
processes having much larger relaxation times go out of equilibrium at this 
point and a temperature dependent chemical potential results for each species so
as to conserve the number corresponding to the measured particle ratios. 
We thus evaluate
the bulk and shear viscosity of a pion gas below the crossover temperature 
in heavy ion collisions 
considering only elastic $\pi\pi$ scattering in the medium with a temperature
dependent pion chemical potential. In the process we 
extend our estimation of the shear viscosity 
at vanishing chemical potential~\cite{Sukanya} where a significant effect of the medium on the
temperature dependence was observed.

In the next section, we briefly recapitulate the expressions for the viscosities
obtained by solving the transport equation in the Chapman-Enskog approximation.
In section III we describe the $\pi\pi$ cross-section in the medium. Numerical
results are provided in section IV followed by summary and conclusions in
section V. Various details concerning the solution of the transport equation are
provided in Appendices A, B and C.  

\section{The bulk and shear viscosity in the Chapman-Enskog approximation}

The evolution
of the phase space distribution of the pions
is governed by the equation
\be
p^\mu\partial_\mu f(x,p)=C[f]
\label{treq}
\ee
where $C[f]$ is the collision integral. For binary elastic 
collisions $p+k\to p'+k'$ which we consider,
this is given by~\cite{Davesne}
\ba
C[f]&=&\int d\Gamma_k\ d\Gamma_{p'}\ d\Gamma_{k'}[f(x,p')f(x,k') \{1+f(x,p)\}
\{1+f(x,k)\}\nonumber\\
&&-f(x,p)f(x,k)\{1+f(x,p')\}\{1+f(x,k')\}]\ W
\ea
where the interaction rate,
\[
W=\frac{s}{2}\ \frac{d\sigma}{d\Omega}(2\pi)^6\delta^4(p+k-p'-k')
\]
and $d\Gamma_q=\frac{d^3q}{(2\pi)^3E}$ with $E=\sqrt{\vec p^2+m_\pi^2}$. 
The $1/2$ factor comes from the indistinguishability of the initial state pions.

For small deviation from local equilibrium we write, in the first Chapman-Enskog
approximation
\be
f(x,p)=f^{(0)}(x,p)+\de f(x,p),~~~\de f(x,p)=f^{(0)}(x,p)[1+f^{(0)}(x,p)]\phi(x,p)
\label{ff}
\ee
where the equilibrium distribution function is given by
\be
f^{(0)}(x,p)=\left[e^{\frac{p^{\mu}u_{\mu}(x)-\mu_\pi(x)}{T(x)}}-1\right]^{-1},
\label{f0}
\ee 
with $T(x)$, $u_\mu(x)$ and $\mu_\pi(x)$ representing the local temperature, 
flow velocity and pion chemical potential respectively. Note that the 
metric 
$diag(1,-1,-1,-1)$ is used. Also, we take $u_\mu u^\mu=1$ where $u_\mu=(1,\vec 0)$
 in the local rest frame.
Putting (\ref{ff}) in (\ref{treq}) 
the deviation function $\phi(x,p)$ is seen to satisfy
\be
p^\mu\partial_\mu f^{(0)}(x,p)=-\cl[\phi]
\label{treq2}
\ee
where the linearized collision term
\ba
\cl[\phi]=&&f^{(0)}(x,p)\int d\Gamma_k\ d\Gamma_{p'}\ d\Gamma_{k'}f^{(0)}(x,k)
\{1+f^{(0)}(x,p')\}\{1+f^{(0)}(x,k')\}\nonumber\\
&&[\phi(x,p)+\phi(x,k)-\phi(x,p')-\phi(x,k')]\ W~.
\ea 
The form of $f^{(0)}(x,p)$ as given in (\ref{f0}) is used on the left side of
(\ref{treq2}) and the time  derivatives are eliminated with the help of equilibrium 
thermodynamic laws. As detailed in Appendix-B this leads us to the equation
\be
[Q\partial_{\nu}u^{\nu}+p_{\mu}\De^{\mu\nu}(p\cdot u- h)
(T^{-1}\partial_{\nu}T-Du_{\nu})-
\langle p_{\mu}p_{\nu} \rangle \langle \del^{\mu} u^{\nu}
\rangle]f^{(0)}(1+f^{(0)})
=-T\cl[\phi]
\label{treq3}
\ee
where $D=u^\mu \partial_\mu$, $\nabla_\mu =\Delta_{\mu\nu} \partial^\nu$,
$\Delta_{\mu\nu}=g_\mn-u_\mu u_\nu$ and the notation
$\la t^\mn\ra\equiv[\frac{1}{2}(\De^{\mu\alpha}\De^{\nu\beta}+
\De^{\nu\alpha}\De^{\mu\beta})-\frac{1}{3}\De^\mn\De^{\alpha\beta}]
t_{\alpha\beta}$ indicates a space-like  
symmetric and traceless form of the tensor $t^\mn$. In this equation
\be
Q=-\frac{1}{3}m_\pi^2+(p\cdot u)^2\{\frac{4}{3}-\gamma'
\}+p\cdot u\{(\gamma''-1) h
-\gamma'''T \}
\ee
where
\be
\gamma'=\frac{(S_{2}^{0}/S_{2}^{1})^2-(S_{3}^{0}/S_{2}^{1})^2+4z^{-1}S_{2}^{0}S_{3}^{1}/(S_{2}^{1})^2+z^{-1}S_{3}^{0}/S_{2}^{1}}
{(S_{2}^{0}/S_{2}^{1})^2-(S_{3}^{0}/S_{2}^{1})^2+3z^{-1}S_{2}^{0}S_{3}^{1}/(S_{2}^{1})^2+2z^{-1}S_{3}^{0}/S_{2}^{1}-z^{-2}}
\ee
\be
\gamma''=1+\frac{z^{-2}}
{(S_{2}^{0}/S_{2}^{1})^2-(S_{3}^{0}/S_{2}^{1})^2+3z^{-1}S_{2}^{0}S_{3}^{1}/(S_{2}^{1})^2+2z^{-1}S_{3}^{0}/S_{2}^{1}-z^{-2}}
\ee
\be
\gamma'''=\frac{S_{2}^{0}/S_{2}^{1}+5z^{-1}S_{3}^{1}/S_{2}^{1}-S_{3}^{0}S_{3}^{1}/(S_{2}^{1})^2}
{(S_{2}^{0}/S_{2}^{1})^2-(S_{3}^{0}/S_{2}^{1})^2+3z^{-1}S_{2}^{0}S_{3}^{1}/(S_{2}^{1})^2+2z^{-1}S_{3}^{0}/S_{2}^{1}-z^{-2}}
\ee
with $z=m_{\pi}/T$ and $h=m_\pi S_{3}^{1}/S_{2}^{1}$. The terms
$S_n^\alpha$ are integrals over Bose functions and are defined in
Appendix-A.
The left hand side of (\ref{treq2}) is thus expressed in terms of the 
thermodynamic forces $\del\cdot u$, $\De^\mn(T^{-1}\partial_{\nu}T-Du_{\nu})$ and
$\langle \del^{\mu} u^{\nu}\rangle$
which have different tensorial ranks, representing a scalar, vector and tensor
respectively. In order to be a solution of this equation
$\phi$ must also be a linear combination of the corresponding
thermodynamic forces.
It is typical to take $\phi$ as
\be
\phi=A\partial\cdot u+B_{\mu}\De^{\mu\nu}(T^{-1}\partial_{\nu}T
-Du_{\nu})-C_{\mu\nu}\langle \partial^{\mu} u^{\nu} \rangle
\label{phi}
\ee
which on 
substitution into (\ref{treq3}) and comparing coefficients of the
(independent) thermodynamic forces on both sides, yields the set of equations
\be
\cl[A]=-Q f^{(0)}(p)\{1+f^{(0)}(p)\}/T
\label{AA}
\ee
\be                             
\cl[C_\mn]=-\la p_\mu p_\nu\ra f^{(0)}(p)\{1+f^{(0)}(p)\}/T
\label{CC}
\ee  
ignoring the equation for $B_\mu$ which is related to thermal conductivity.
These integral equations are to be solved to get the coefficients
$A$ and $C_\mn$. It now remains to link these to the viscous coefficients
$\zeta$ and $\eta$. This is achieved by means of the dissipative part of the 
energy-momentum tensor resulting from the use of the non-equilibrium 
distribution function (\ref{ff}) in
\be
T^\mn=\int d\Gamma_p\ p^\mu p^\nu f(p)=T^{\mn (0)}+\De T^{\mn}
\ee
where
\be
\Delta T^{\mu\nu}=\int d\Gamma_p f^{(0)} (1+f^{(0)}) C_{\alpha\beta} \langle
p^{\alpha}p^{\beta} 
\rangle \langle \partial^{\mu} u^{\nu} \rangle
+\int d\Gamma_p f^{(0)} (1+f^{(0)} )QA\Delta^{\mu\nu}\del\cdot u
\ee
Again, for a small deviation $\phi(x,p)$, close to equilibrium, so that
only first order derivatives contribute, the dissipative tensor can be
generally expressed in the form~\cite{Purnendu,Polak}
\be
\Delta T^{\mu\nu}=-2\eta\langle \partial^{\mu} u^{\nu} \rangle
-\zeta\Delta^{\mu\nu}\partial\cdot u~.
\ee
Comparing,
we obtain the expressions of shear and bulk viscosity,
\be
\eta=-\frac{1}{10}\int d\Gamma_p\ C_\mn\la p^\mu p^\nu \ra
f^{(0)}(p)\{1+f^{(0)}(p)\}
\ee
and
\be
\zeta=-\int d\Gamma_p \ QA f^{(0)}(p)\{1+f^{(0)}(p)\}~.
\ee
The coefficients $A$ and $C_{\mu\nu}$ are perturbatively obtained from
(\ref{AA}) and (\ref{CC})
by expanding in terms of orthogonal polynomials which reduces the integral
equations to algebraic ones. 
This elaborate procedure using Laguerre polynomials, is described
in~\cite{Polak,Davesne}. A brief account is provided in Appendix C.
Finally, the first 
approximation to shear and bulk viscosity come out as
\be
\eta=\frac{T}{10}\ \frac{\gamma_0^2}{c_{00}}
\label{eta}
\ee
and
\be
\zeta=T\frac{\alpha_{2}^{2}}{a_{22}}
\label{zeta}
\ee
where
\ba
\gamma_0&=&-10\frac{S_3^{2}(z)}{S_2^{1}(z)}~,\nonumber\\
c_{00}&=&16\{I_1(z)+I_2(z)+\frac{1}{3}I_3(z)\}~,
\ea
and
\ba
\alpha_{2}&=&\frac{z^3}{2} [ \frac{1}{3}(\frac{S_{3}^0}{S_{2}^{1}}-z^{-1})
+(\frac{S_{2}^{0}}{S_{2}^{1}}+\frac{3}{z}\frac{S_{3}^{1}}{S_{2}^{1}})
\{(1-\gamma'')\frac{S_{3}^{1}}{S_{2}^{1}}+\gamma'''z^{-1}\}
\nonumber\\
&-&(\frac{4}{3}-\gamma')\{ \frac{S_{3}^{0}}
{S_{2}^{1}}+15z^{-2}\frac{S_{3}^{2}}{S_{2}^{1}}+2z^{-1}\}]~,
\nonumber\\a_{22}&=&2z^2I_{3}(z)~.
\ea

The integrals $I_\alpha(z)$ appearing in the above expressions are defined as
\ba
I_\alpha(z)&=&\frac{z^4}{[S_2^{1}(z)]^2} \ e^{(-2\mu_\pi/T)}\int_0^\infty d\psi\ \cosh^3\psi
\sinh^7\psi\int_0^\pi
d\Theta\sin\Theta\frac{1}{2}\frac{d\sigma}{d\Omega}(\psi,\Theta)\int_0^{2\pi}
d\phi\nonumber\\&&\int_0^\infty d\chi \sinh^{2\alpha} \chi
\int_0^\pi d\theta\sin\theta\frac{e^{2z\cosh\psi\cosh\chi}}
{(e^E-1)(e^F-1)(e^G-1)(e^H-1)}\ M_\alpha(\theta,\Theta)
\label{c00_bose}
\ea
in which the functions $M_\alpha(\theta,\Theta)$ represent
\ba
M_1(\theta,\Theta)&=&1-\cos^2\Theta~,\nonumber\\M_2(\theta,\Theta)&=&\cos^2\theta+\cos^2\theta'
-2\cos\theta\cos\theta'\cos\Theta~,\nonumber\\
M_3(\theta,\Theta)&=&[\cos^2\theta-\cos^2\theta']^2~.
\ea

Note that the differential cross-section which appears in the denominator
is the dynamical input in the expressions for $\eta$ and $\zeta$. It is this
quantity we turn to in the next section. 

\section{The $\pi\pi$ cross-section with medium effects} 

\begin{figure}
\includegraphics[scale=0.5]{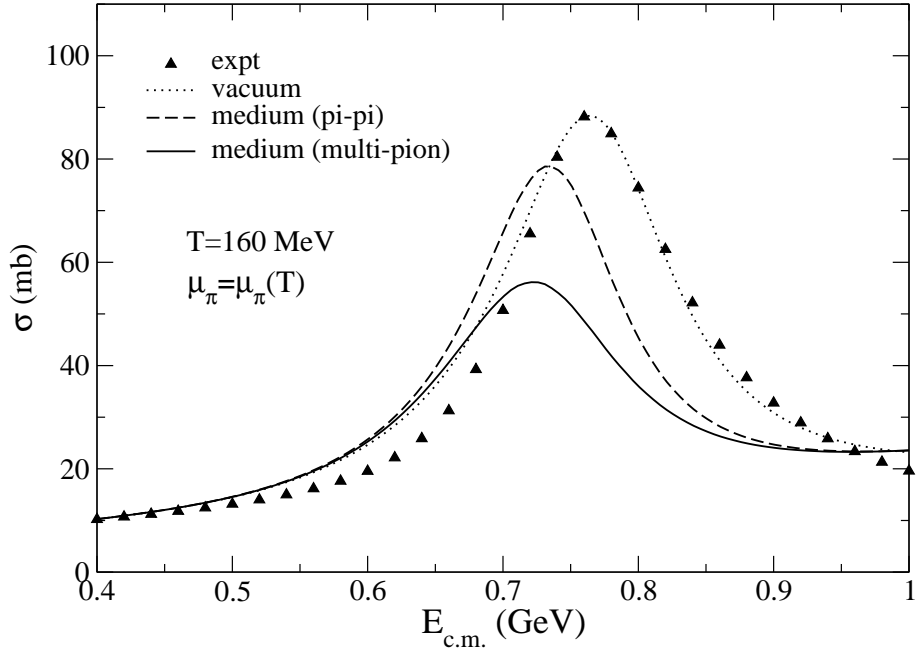}
\caption{The $\pi\pi$ cross-section as a function of centre of mass energy. The
dotted line indicates the cross-section obtained using eq.~(\ref{amp}) which
agrees well with the experimental values (eq.~(\ref{expt})) shown by filled
triangles. The dashed
and solid lines depict the in-medium cross-section for $\pi\pi$ and 
$\pi h~(=\pi,\om,h_1,a_1)$ 
loops respectively in the $\rho$ and $\sigma$ self-energies evaluated at $T$=160 MeV
with temperature dependent pion chemical potential.}
\label{sigmafig}
\end{figure}

\begin{figure}
\includegraphics[scale=0.5]{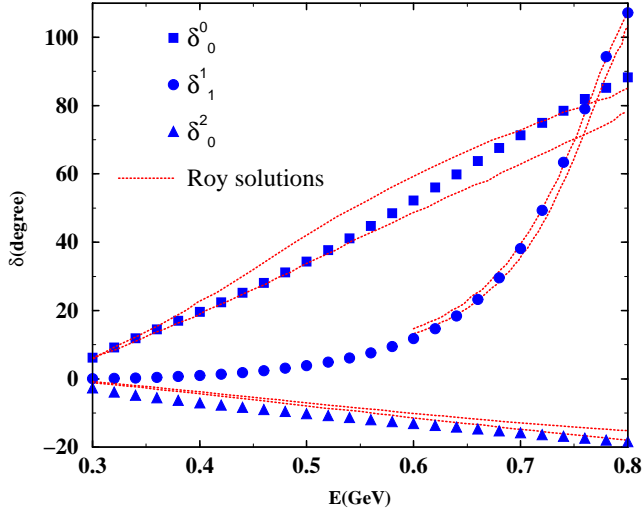}
\caption{(Color online) The $\pi\pi$ phaseshifts (\ref{phaseshifts}) as a function of $E$ compared
to Roy equation solutions given in Ref.~\cite{Anantha}}
\label{Royfig}
\end{figure}

The strong interaction dynamics of the pions enters the collision integrals
through the cross-section. In Fig.~\ref{sigmafig} we show the $\pi\pi$
cross-section as a function of the centre of mass energy of scattering. The
different curves are explained below.
The filled triangles referred to as experiment is a widely used resonance saturation
parametrization~\cite{Bertsch1,Welke} of isoscalar and isovector phase shifts obtained from various
empirical data involving the $\pi\pi$ system. The isospin averaged 
differential cross-section
is given by
\be
\frac{d\sigma(s)}{d\Omega}=\frac{4}{q_{cm}^2}\left[\frac{1}{9}
\sin^2\de^0_0+\frac{5}{9}\sin^2\de^2_0+\frac{1}{3}\cdot
9\sin^2\de_1^1\cos^2\theta\right]
\label{expt}
\ee
where
\ba
\de^0_0&=&\frac{\pi}{2}+\arctan\left(\frac{E-m_\sigma}{\Gamma_\sigma/2}\right)\nonumber\\
\de_1^1&=&\frac{\pi}{2}+\arctan\left(\frac{E-m_\rho}{\Gamma_\rho/2}\right)\nonumber\\
\de^2_0&=&-0.12p/m_\pi~.
\label{phaseshifts}
\ea
The widths are given by $\Gamma_\sigma=2.06p$ and 
$\Gamma_\rho=0.095p\left(\frac{p/m_\pi}{1+(p/m_\rho)^2}\right)^2$
with $m_\sigma=5.8m_\pi$ and $m_\rho=5.53m_\pi$~. As seen in Fig.~\ref{Royfig}
these phase shifts agree quite well with those obtained from solutions of the Roy
equations as given in~\cite{Anantha}. The bands bordered by the dotted lines
represent the uncertainties in the solution. The experimentally measured
phase shifts (not shown) have error bars~\cite{Anantha} which are not reflected
in the parametrizations (\ref{phaseshifts}) plotted in Fig.~\ref{sigmafig}.

Our objective now is to set up a dynamical model which agrees 
reasonably with the above parametrization in vacuum and
is amenable to the
incorporation of medium effects using many body techniques.
We thus evaluate the $\pi\pi$ cross-section
involving $\rho$ and $\sigma$ meson exchange processes using the interaction
Lagrangian
\be
\cl=g_\rho\vec\rho^\mu\cdot\vec \pi\times\del_\mu\vec \pi+\frac{1}{2}g_\sigma
m_\sigma\vec \pi\cdot\vec\pi\sigma
\ee
where $g_\rho=6.05$ and $g_\sigma=2.5$.
In the matrix elements corresponding to $s$-channel $\rho$ and $\sigma$
exchange diagrams which appear for total isospin $I=1$ and 0 respectively, we
introduce a decay width in the corresponding propagator. We get~\cite{Sukanya},
\ba
\cm_{I=0}&=&
g_\sigma^2 m_\sigma^2\left[\frac{3}{s-m_\sg^2+im_\sg\Gm_\sg}+\frac{1}{t-m_\sg^2}
+\frac{1}{u-m_\sg^2}\right]
+2g_\rho^2\left[\frac{s-u}{t-m_\rho^2}+\frac{s-t}{u-m_\rho^2}\right]
\nonumber\\
\cm_{I=1}&=&g_\sigma^2 m_\sigma^2\left[\frac{1}{t-m_\sg^2}-\frac{1}{u-m_\sg^2}\right]
+g_\rho^2\left[\frac{2(t-u)}{s-m_\rho^2+im_\rho\Gamma_\rho}+
\frac{t-s}{u-m_\rho^2}-\frac{u-s}{t-m_\rho^2}\right]
\nonumber\\
\cm_{I=2}&=&g_\sigma^2 m_\sigma^2\left[\frac{1}{t-m_\sg^2}+\frac{1}{u-m_\sg^2}\right]
+g_\rho^2\left[\frac{u-s}{t-m_\rho^2}+\frac{t-s}{u-m_\rho^2}\right]~.
\label{amp}
\ea
The differential cross-section is then obtained from 
$\frac{d\sigma}{d\Omega}=\overline{|\cm|^2}/64\pi^2 s$ 
where the isospin averaged amplitude is given by
$\overline{|\cm|^2}=\frac{1}{9}\sum(2I+1)\overline{|\cm_I|^2}$.
 
Considering the isoscalar and isovector contributions which involves 
$s$-channel $\sigma$ and $\rho$ meson exchange respectively, 
the integrated cross-section, 
shown by the dotted line (indicated by 'vacuum') in Fig.~\ref{sigmafig} 
is seen to agree 
reasonably well with the experimental cross-section 
up to a centre of mass energy of about 1 GeV beyond which the theoretical
estimate gives higher values. We hence use the experimental cross-section beyond
this energy. Inclusion of the non-resonant $I=2$ contribution results in
 an overestimation and 
is ignored in this normalization of the model to experimental data. 
It is essential at this  point to emphasize that the present approach
of introducing finite decay widths for the exchanged resonances explicitly
is not consistent with the well-known chiral 
low energy theorems concerning the amplitude for $\pi\pi$
scattering~\cite{Weinberg_PRL} as well as the scattering lengths.

We now construct the in-medium cross-section by introducing
the effective propagator for the $\rho$ and $\sigma$ mesons 
in the above expressions for the matrix
elements. We first discuss the case of the $\rho$ followed by the 
relatively simpler case of the $\sigma$ meson.

\begin{figure}
\includegraphics[scale=0.5]{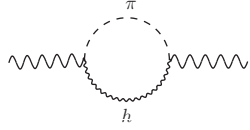}
\caption{Self-energy diagrams where $h$ stands for
$\pi,\om,h_1,a_1$ mesons.}
\label{selfdiag}
\end{figure}

The in-medium propagator of the $\rho$ is obtained in terms of the 
self-energy by solving the Dyson equation and is given by
\be
D_\mn=D^{(0)}_\mn+D^{(0)}_{\mu\sigma}\Pi^{\sigma\lambda}D_{\lambda\nu}
\label{eq:dyson}
\ee
where $D^{(0)}_\mn$ is the vacuum propagator for the $\rho$ meson
and $\Pi^{\sigma\lambda}$ is the self energy function obtained
from one-loop diagrams shown in Fig.~\ref{selfdiag}.
The standard procedure~\cite{Bellac} to solve this equation in the medium is to 
decompose the self-energy into transverse and longitudinal components.
For the case at hand the difference between these components 
is found to be small and is hence ignored.
We work with the polarization averaged self-energy function defined as
\be
\Pi=\frac{1}{3}(2\Pi^T+q^2\Pi^L)
\ee
where
\be
\Pi^T=-\frac{1}{2}(\Pi_\mu^\mu +\frac{q^2}{\bar q^2}\Pi_{00}),~~~~
\Pi^L=\frac{1}{\bar q^2}\Pi_{00} , ~~~\Pi_{00}\equiv u^\mu u^\nu \Pi_{\mn}~.
\label{pitpil}
\ee
The in-medium propagator is then written as
\be
D_\mn(q_0,\vec q)=\frac{-g_\mn+q_\mu q_\nu/q^2}{q^2-m_\rho^2-{\rm Re}\Pi(q_0,\vec q)+
i{\rm Im}\Pi(q_0,\vec q)}~.
\label{medprop}
\ee
The scattering, decay and regeneration processes which cause a gain or loss of
$\rho$ mesons in the medium are responsible for the imaginary part of its 
self-energy. 
The real part on the other hand modifies the position
of the pole of the spectral function.

In the real-time formulation of thermal field theory the self-energy
assumes a 2$\times$2 matrix structure of which the 11-component is given by 
\be
\Pi_{\mn}^{11}(q)=i\int\frac{d^4k}{(2\pi)^4}N_{\mn}(q,k)D_\pi ^{11}(k)D_h^{11}(q-k)
\ee
where $D^{11}$ is the 11-component of the scalar propagator given by
$D^{11}(k)=\De(k)+2\pi if^{(0)}(k)\de(k^2-m^2)$. It turns out that the 
real and imaginary parts of the self-energy
function which appear in eq.~(\ref{medprop}) can be obtained in terms of the 11-component through the
relations~\cite{Bellac,Mallik_RT}
\ba
{\rm Re}\,\Pi_{\mn}&=&{\rm Re}\,\Pi_{\mn}^{11}\nonumber\\
{\rm Im}\Pi_{\mn}&=&\epsilon(q_0)\tanh(\beta q_0/2){\rm Im}\,\Pi_{\mn}^{11}~.
\ea
Tensor structures associated with the two  
vertices and the vector propagator are included in $N_{\mn}$ and are
available in~\cite{Ghosh1} where the interactions were
taken from chiral perturbation theory.
It is easy to perform the integral over $k_0$ using suitable contours to obtain
\ba
{\Pi}^{\mn}(q_0,\vq)&=&\int\frac{d^3k}{(2\pi)^3}\frac{1}
{4\om_{\pi}\om_{h}}\left[\frac{(1+f^{(0)}(\omp))N^{\mn}_1+f^{(0)}(\omh)N^{\mn}_3}
{q_0 -\om_{\pi}-\om_{h}+i\eta\ep(q_0)}
+\frac{-f^{(0)}(\omp)N^{\mn}_1+f^{(0)}(\omh)N^{\mn}_4}
{q_0-\om_{\pi}+\om_{h}+i\eta\ep(q_0)} 
\right.\nonumber\\
&&+\left. \frac{f^{(0)}(\omp)N^{\mn}_2 -f^{(0)}(\omh)N^{\mn}_3}
{q_0 +\om_{\pi}-\om_{h}+i\eta\ep(q_0)} 
+\frac{-f^{(0)}(\omp)N^{\mn}_2 -(1+f^{(0)}(\omh))N^{\mn}_4}
{q_0 +\om_{\pi}+\om_{h}+i\eta\ep(q_0)}\right]
\label{MM_rho}
\ea
where $f^{(0)}(\om)=\frac{1}{e^{(\om-\mu_\pi)/T}-1}$ is the Bose distribution
function with arguments $\om_\pi=\sqrt{\vk^2+m_\pi^2}$ and
$\om_h=\sqrt{(\vq-\vk)^2+m_h^2}$. Note that this expression is a generalized
form for the in-medium self-energy obtained by Weldon~\cite{Weldon}.
The subscript $i(=1,..4)$ on $N^{\mn}$ in (\ref{MM_rho}) correspond to its values for 
$k_0=\om_\pi,-\om_\pi,q_0-\om_h,q_0+\om_h$ respectively. It is easy to read off
the real and imaginary parts from (\ref{MM_rho}). The angular integration can be
carried out using the $\de$-functions in each of the four terms in the imaginary
part which define the kinematically allowed regions in $q_0$ and $\vq$ where
scattering, decay and regeneration processes occur in the medium leading to the
loss or gain of $\rho$ mesons~\cite{Ghosh1}. 
The vector mesons $\omega$, $h_1$ and $a_1$ which appear in the loop
have negative
$G$-parity and have substantial $3\pi$ and $\rho\pi$ decay widths~\cite{PDG}. The
(polarization averaged) self-energies containing these unstable particles in the loop graphs have thus been folded 
with their spectral functions,
\be
\Pi(q,m_h)= \frac{1}{N_h}\int^{(m_h+2\Gm_h)^2}_{(m_h-2\Gm_h)^2}dM^2\frac{1}{\pi} 
{\rm Im} \left[\frac{1}{M^2-m_h^2 + iM\Gm_h(M) } \right] \Pi(q,M) 
\ee
with $N_h=\displaystyle\int^{(m_h+2\Gm_h)^2}_{(m_h-2\Gm_h)^2}
dM^2\frac{1}{\pi} {\rm Im}\left[\frac{1}{M^2-m_h^2 + iM\Gm_h(M)} \right]$.
The contributions from the
loops with heavy mesons (the $\pi h$ loops) may then be considered as a multi-pion contribution to the
$\rho$ self-energy. 

The medium effect on propagation of the $\sigma$ meson is estimated analogously as
above. The effective propagator in this case is given by
\be
D(q_0,\vec q)=\frac{-1}{q^2-m_\sg^2-{\rm Re}\Pi(q_0,\vec q)+
i{\rm Im}\Pi(q_0,\vec q)}~.
\label{medpropsig}
\ee 
Following the steps outlined above the expression for the self-energy
of the $\sigma$ is given by
\ba
{\Pi}(q_0,\vq)&=&N\int\frac{d^3k}{(2\pi)^3}\frac{1}
{4\om_{\pi}\om_{\pi}'}\left[\frac{1+f^{(0)}(\omp)+f^{(0)}(\omp')}
{q_0 -\om_{\pi}-\om_{\pi}'+i\eta\ep(q_0)}
+\frac{f^{(0)}(\omp')-f^{(0)}(\omp)}
{q_0-\om_{\pi}+\om_{\pi}'+i\eta\ep(q_0)} 
\right.\nonumber\\
&&+\left. \frac{f^{(0)}(\omp)-f^{(0)}(\omp')}
{q_0 +\om_{\pi}-\om_{\pi}'+i\eta\ep(q_0)} 
-\frac{1+f^{(0)}(\omp)+f^{(0)}(\omp')}
{q_0 +\om_{\pi}+\om_{\pi}'+i\eta\ep(q_0)}\right]
\label{MM_sig}
\ea
where $\omp'=\sqrt{(\vq-\vk)^2+m_\pi^2}$.
The imaginary part for the kinematic region of our interest in this case receives
contribution only from the first term which essentially describes the decay of
the $\sigma$ into two pions minus the reverse process of formation.
 
The in-medium cross-section is now obtained by using the full $\rho$
and $\sg$ propagators given by
(\ref{medprop}) and (\ref{medpropsig}) respectively in place of the vacuum propagators in the 
scattering amplitudes. The long dashed line in Fig.~\ref{sigmafig} shows a 
suppression of the peak when only the $\pi\pi$ loop in the $\rho$ and $\sigma$ 
self-energies are considered. This
effect is magnified when the $\pi h$ loops containing heavier mesons in the
$\rho$ self-energy are taken into account and is depicted by the
solid line indicated by multi-pion. 
This is also accompanied by a small shift in the peak position.
Extension to the case of finite baryon density can be done 
using the spectral function computed in~\cite{Ghosh2} where an extensive
list of baryon (and
anti-baryon) loops are considered along with the mesons.
A similar modification of the $\pi\pi$ cross-section for a hot and dense
system was seen also in~\cite{Bertsch2}.

\begin{figure}
\includegraphics[scale=0.5]{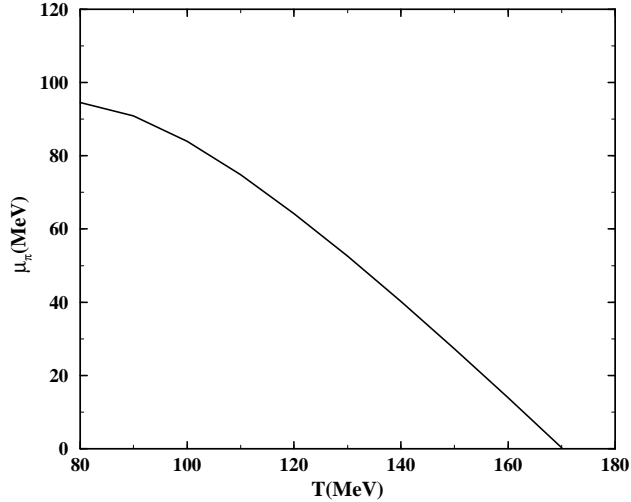}
\caption{The pion chemical potential as a function of temperature~\cite{Hirano}.}
\label{Hiranofig}
\end{figure}

We end this section with a discussion of the pion chemical potential. As mentioned in
the introduction, in heavy ion collisions pions get out of chemical equilibrium
early at $T\sim$ 170 MeV and a corresponding chemical potential starts building up
with decrease in temperature. The kinetics of the gas is then dominated by 
elastic collisions including resonance formation such as 
$\pi\pi\leftrightarrow\rho$ etc. At still lower temperature, $T\sim$ 100 MeV
elastic collisions become rarer and the momentum distribution gets frozen
resulting in kinetic freeze-out. This scenario is quite compatible with the
treatment of medium modification of the $\pi\pi$ cross-section being employed
in this work where the $\pi\pi$ interaction is mediated by $\rho$ and $\sigma$
exchange and the subsequent propagation of these mesons are modified by two-pion
and effective multi-pion fluctuations. 
We take the temperature dependent pion chemical potential from 
Ref.~\cite{Hirano} which implements the formalism described in~\cite{Bebie}
and reproduces the slope of the transverse momentum spectra of identified
hadrons observed in experiments.
Here, by
fixing the ratio $s/n$ where $s$ is the entropy density and $n$ the number
density to the value at chemical freeze-out where $\mu_\pi=0$,
one can go down in
temperature up to the kinetic freeze-out by increasing the pion chemical
potential. This provides the temperature dependence leading to $\mu_\pi(T)$
which is shown in Fig.~\ref{Hiranofig}.
In this partial chemical equilibrium scenario of~\cite{Bebie}
the chemical potentials of the heavy mesons are determined from elementary
processes. The $\om$ chemical potential e.g. is given  by
$\mu_\om=3\times0.88\mu_\pi$,
 as a consequence of the processes 
$\om\leftrightarrow\pi\pi\pi$ occurring in the medium.
The branching ratios are taken from~\cite{PDG}.

\section{Results}

\begin{figure}
\includegraphics[scale=0.5]{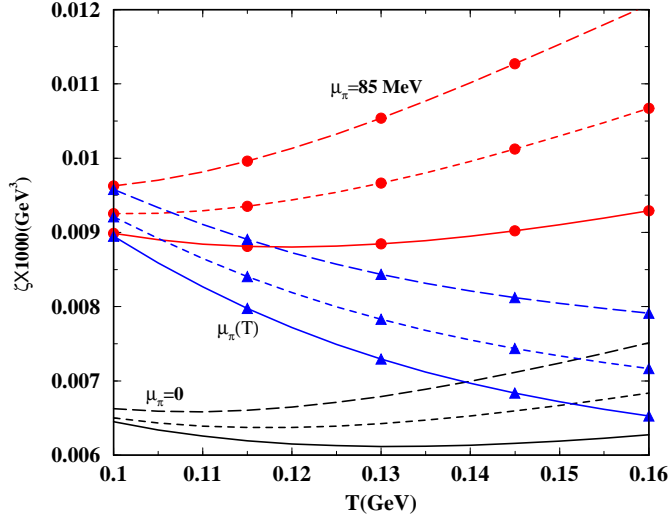}
\caption{(Color online)The bulk viscosity in various scenarios as a function of $T$. 
The upper (with circles), middle (with triangles) and lower sets of curves correspond to $\mu_\pi=85$ MeV,
$\mu_\pi=\mu_\pi(T)$ and $\mu_\pi=0$ respectively. In each set the solid line
represents use of vacuum cross-section, the dotted line for in-medium
modification due to pion loop 
and the dashed line for loops with heavy particles in addition.}
\label{zetafig}
\end{figure}

We begin with the results for bulk viscosity $\zeta$ as a function of 
temperature $T$. In Fig.~\ref{zetafig} the three sets of curves
correspond to different values of the pion chemical potential. 
The uppermost set of curves (with circles) show the bulk viscosity 
calculated with a pion chemical
potential $\mu_\pi\sim 85$ MeV. The corresponding curves in the lowermost set 
are evaluated with $\mu_\pi=0$. These values are representative
of the kinetic and chemical freeze-out in heavy ion collisions respectively.
 The solid line
in the lowermost set represents the case where the vacuum cross-section 
given by eq.~(\ref{amp}) is used and agrees with the estimate
in~\cite{Davesne}. 
The set of curves with triangles depicts 
the situation when $\mu_\pi$ is a
(decreasing) function of temperature as shown in Fig.~\ref{Hiranofig}. This resembles
the situation encountered in the later stages of heavy ion collisions and
interpolates between the results with the constant values of the pion chemical
potential discussed before. The three
curves in each set show the effect of medium on the $\pi\pi$ cross-section.
The short-dashed lines in each of the sets depict medium effects for pion 
loops in the $\rho$ propagator and the long dashed lines correspond to
the situation when the heavy mesons are included i.e. for 
$\pi h$ loops where $h=\pi,\om,h_1,a_1$. The clear separation between the
curves in each set displays a significant effect brought about
by the medium dependence of the cross-section. A
large dependence on the pion chemical potential is also inferred
since the three sets of curves appear nicely separated.
\begin{figure}
\includegraphics[scale=0.5]{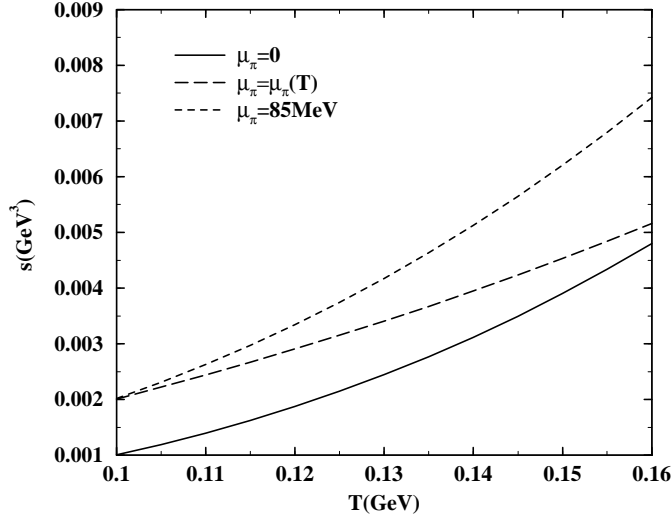}
\caption{The entropy density of an interacting pion gas as a function of $T$ for different values of the pion
chemical potential.}
\label{entropyfig}
\end{figure}

\begin{figure}
\includegraphics[scale=0.5]{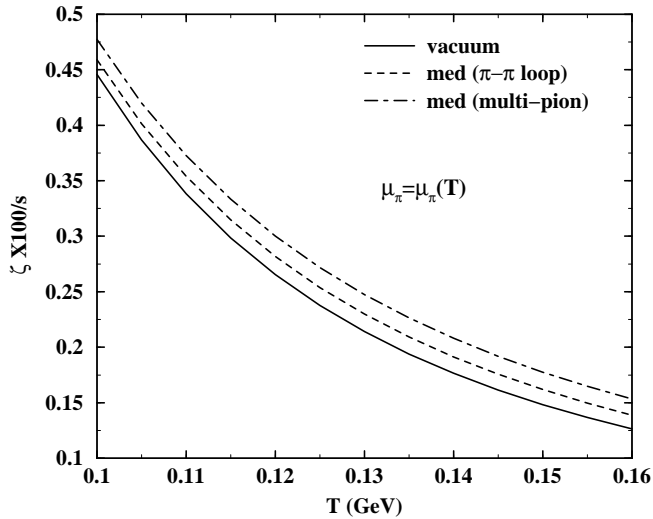}
\caption{$\zeta/s$ as a function of $T$ for different $\pi\pi$ cross-section. The
temperature dependent pion chemical potential has been used in all cases.}
\label{zetabysfig}
\end{figure} 

Viscosities for relativistic fluids are generally expressed in terms of a
dimensionless ratio obtained by dividing with the entropy density. 
The latter
is obtained from the thermodynamic relation
\be
Ts=\ep +P -n\mu_\pi~.
\ee
For a free pion gas, we get on using the relations for the energy density $\ep$,
pressure $P$ and number density $n$ from Appendix-A,
\be
s=\frac{g_\pi}{2\pi^2}\ m_\pi^2[m_\pi S_3^{1}(z)-\mu_\pi S_2^{1}(z)]~.
\ee
Here $g_\pi=3$, $z=m_\pi/T$ and the functions $S_n^\alpha(z)$ are given
in Appendix-A
. Interactions between pions lead to corrections to this formula.
To $O(T^6)$ this has been calculated for finite pion chemical potential
in~\cite{Nicola} using chiral perturbation theory to give
\be
\De s=-\frac{3m_\pi^4}{16\pi^4 f_\pi^2}S_1^1(z)[m_\pi S_2^0(z)-\mu_\pi S_1^0(z)]
\ee
where $f_\pi=93$ MeV.
It is easily verified that this expression reduces for $\mu_\pi=0$ to those
given in~\cite{Gerber,Lang}. This correction is
$\sim 1-2\%$ for values of $\mu_\pi$ and $T$ considered here.
In Fig.~\ref{entropyfig} the entropy density of an interacting pion gas
as a function of temperature is
shown for three values of the pion chemical potential as discussed in the
context of Fig.~\ref{zetafig}.

In Fig.~\ref{zetabysfig} we show $\zeta/s$ as a function of $T$ using the 
temperature dependent pion chemical 
potential. The medium dependence is clearly observed when we compare
the results obtained with the vacuum cross-section with the ones where the
$\sigma$ and $\rho$ propagation is modified due to $\pi\pi$ and $\pi h$ 
(multi-pion) loops. 
The decreasing trend with increasing temperature was observed also
in~\cite{Moore} and~\cite{Dobado_bulk}.

\begin{figure}
\includegraphics[scale=0.5]{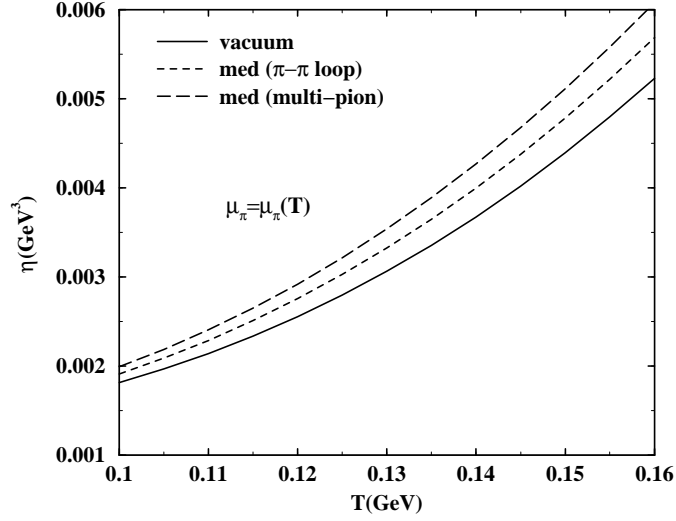}
\caption{$\eta$ as a function of $T$ for different values of the $\pi\pi$
cross-section. The
temperature dependent pion chemical potential has been used in all cases.}
\label{etafig}
\end{figure}

\begin{figure}
\includegraphics[scale=0.5]{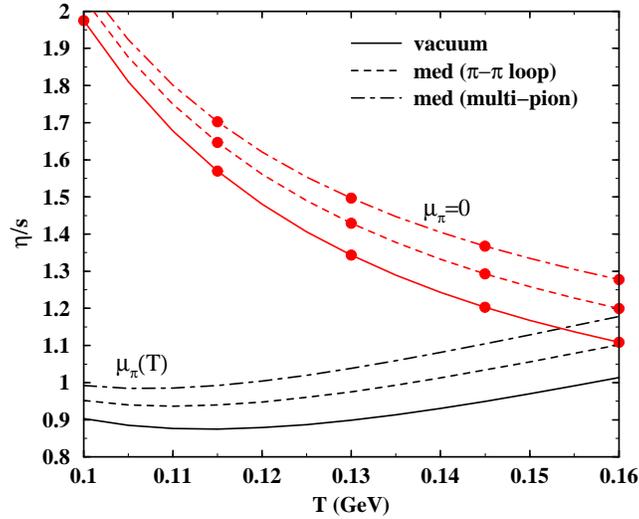}
\caption{(Color online)$\eta/s$ as a function of $T$. The upper set of curves with
circles correspond to $\mu_\pi=0$ and the lower set corresponds to
$\mu_\pi=\mu_\pi(T)$.}
\label{etabysfig}
\end{figure} 

We now turn to the shear viscosity. Here we extend the work in Ref.~\cite{Sukanya}
to include the effect of the pion chemical potential. Shown in Fig.~\ref{etafig}
is the shear viscosity as a function of $T$ where the results with $\pi\pi$ and
$\pi h$ loops are contrasted with the case where the vacuum cross-section is
used. The result with the vacuum cross-section agrees
with~\cite{Dobado3} and~\cite{Davesne} for $\mu_\pi=0$.
A noticeable medium effect is observed as indicated by the short and long-dashed
lines. 

Finally, in Fig.~\ref{etabysfig} we show results
for $\eta/s$. For $\mu_\pi=0$ the upper set of curves with filled circles show the
usual decreasing trend as seen, for example in~\cite{Nakano}. This trend is reversed
when $\mu_\pi(T)$ is used and $\eta/s$
increases with $T$  in contrast to $\zeta/s$ which
decreases. The values in all cases remain well above $1/4\pi$.

\section{Summary and Outlook}

In this work we have evaluated the viscosities of a pion gas by solving the
Uehling-Uhlenbeck transport equation in the Chapman-Enskog approximation to
leading order with an aim to study the effect of a medium dependent
cross-section. The $\pi\pi$ cross-section which goes as an input to these
calculations is evaluated from $\sigma$ and $\rho$ exchange processes. 
Spectral
modifications of the $\rho$ and $\sigma$ propagators implemented through 
$\pi h$ and $\pi \pi$
self-energy diagrams respectively show a significant effect on the cross-section
and consequently on the temperature dependence of
transport coefficients. The effect of early chemical freeze-out in heavy ion
collisions is implemented through a temperature dependent pion chemical potential
which also enhances the medium effect. Results for $\zeta/s$
and $\eta/s$ also show a significant medium dependence in this scenario. 

The viscous coefficients and their temperature dependence could affect
the quantitative estimates of signals of heavy ion collisions particularly
where hydrodynamic simulations are involved.
For example, it has been argued in~\cite{Dusling} that corrections to the freeze-out distribution
due to bulk viscosity can be significant. As a result the hydrodynamic 
description of the $p_T$ spectra and elliptic flow of hadrons could be improved
by including a realistic temperature dependence of the viscous coefficients.
Studies in this direction are in progress. 

\section{Appendix A}
\setcounter{equation}{0}
\renewcommand{\theequation}{A.\arabic{equation}}

The integrals over Bose functions which appear in the definitions of 
thermodynamic quantities like energy density, pressure, entropy etc. 
of a pion gas can be expressed in terms
of the functions $S_n^\alpha(z)$ where $z=m/T$. 
Some of these which appear
in various expressions in this work are given by
\ba
S_1^1(z)&=&\frac{2\pi^2}{zT^2}\int\ d\Gm_p f^{(0)}(p)\nn\\
S_2^1(z)&=&\frac{2\pi^2}{z^2T^3}\int\ d\Gm_p\ E\  f^{(0)}(p)\nn\\
S_3^2(z)&=&\frac{2\pi^2}{3z^3T^5}\int\ d\Gm_p\ p^2E\  f^{(0)}(p)
\ea
where $E=\sqrt{p^2+m^2}$, $f^{(0)}(p)=[e^{\beta(E-\mu)}-1]^{-1}$ and
$d\Gm_p=d^3p/(2\pi)^3E$. Using the formula
$[a-1]^{-1}=\displaystyle\sum_{n=1}^\infty(a^{-1})^n$
these integrals can be converted to sums over infinite series and
can be compactly expressed as
$S_n^\alpha(z)=\displaystyle\sum_{k=1}^\infty e^{k\mu/T} k^{-\alpha} K_n(kz)$, 
$K_n(x)$ denoting the modified Bessel function of order $n$ given by
\be
K_n(x)=\frac{2^n n!}{(2n)!\ x^n}\int_x^\infty\ d\tau
(\tau^2-x^2)^{n-\frac{1}{2}}e^{-\tau}
\ee
or
\be
K_n(x)=\frac{2^n n!(2n-1)}{(2n)!\ x^n}\int_x^\infty\ \tau d\tau
(\tau^2-x^2)^{n-\frac{3}{2}}e^{-\tau}~.
\ee
Using now the property
\be
\frac{\partial}{\partial(\mu/T)}S_n^\alpha(z)=S_n^{\alpha-1}(z)
\ee
the remaining integrals may be easily obtained. 
The equilibrium formulae for the number density, pressure, energy density and
enthalpy density are respectively given by 
\ba
n&=&g\int d\Gamma_p E_p
f^{(0)}(p)=\frac{g}{2\pi^2}z^2T^3S_2^1(z),\nn\\
P&=&g\int
d\Gamma_p\frac{\vp^2}{3}f^{(0)}(p)=\frac{g}{2\pi^2}z^2T^4S_2^2(z),\nn\\
\ep=ne&=&g\int d\Gamma_pE_p^2f^{(0)}(p)
=\frac{g}{2\pi^2}z^2T^4[z S_3^1(z)-S_2^2(z)]\nn\\
H&=&nh=nzT\frac{S_3^1(z)}{S_2^1(z)}~,
\ea
where for a pion gas $g=3$. In these appendices we have suppressed the subscript
'$\pi$' on $m$ and $\mu$ for brevity.
\section{Appendix B}
\setcounter{equation}{0}
\renewcommand{\theequation}{B.\arabic{equation}}

Here we show how the left side of the linearized transport 
equation (\ref{treq2}) given by
\be
p^\mu\partial_\mu f^{(0)}(x,p)=-\cl[\phi]
\ee
can be expressed in terms of thermodynamic forces. We write $\del_\mu=u_\mu
u^\nu\del_\nu+\De_\mn\del^\nu\equiv u_\mu D+\nabla_\mu$ separating the time derivative $D$ 
and the gradient $\nabla_\mu$. Note that $D\to\del/\del t$ and
$\nabla_\mu\to\del_i$ in the local rest frame. On differentiating $f^{(0)}$ 
\be
(p\cdot u)\left[\frac{p\cdot u}{T^2}DT+D(\frac{\mu}{T})-\frac{p^\mu}{T}
Du_\mu\right]+p^\mu\left[\frac{p\cdot
u}{T^2}\nabla_\mu T+\nabla_\mu(\frac{\mu}{T})
-\frac{p^\nu}{T}\nabla_\mu u_\nu\right]=-\frac{\cl[\phi]}{f^{(0)}(1+f^{(0)})}~.
\label{tr_long}
\ee
The terms $DT$ and $D(\frac{\mu}{T})$ do not appear in the expression of the
thermodynamic forces and are to be eliminated using equilibrium laws. From the
equation of continuity
\[\del_\mu N^\mu=0\]
where $N^\mu=nu^\mu$,  we get
\be
Dn=-n\del_\mu u^\mu~.
\ee
Again, contracting the equation for energy-momentum conservation with $u_\mu$
i.e.
\[u_\mu\del_\nu T^{\mn\, (0)}=0\]
where $T^{\mn\, (0)}=n[(e+\frac{P}{n})u^\mu u^\nu-g^\mn \frac{P}{n}]$ results in the relation
\be
De=-\frac{P}{n}\del_\mu u^\mu~.
\label{De}
\ee
Further, contracting with the projector $\De^\mn$,
\[\De^\mu_\nu\del_\alpha T^{\nu\alpha\, (0)}=0\]
yields 
\be
D u^\mu=\frac{1}{nh}\nabla^\mu P~.
\label{Du}
\ee
Making use of the relativistic Gibbs-Duhem relation
\[\del^\mu P=nT\del^\mu(\frac{\mu}{T})+nhT^{-1}\del^\mu T\]
and (\ref{De}) in (\ref{Du}) we get
\be
Dh=TD(\frac{\mu}{T})+hT^{-1}DT~.
\label{Dh}
\ee
Using now the expansions
\ba
De&=&\left.\frac{\del e}{\del T}\right|_{\mu/T}DT+
\left.\frac{\del e}{\del(\mu/T)}\right|_TD(\frac{\mu}{T})\nn\\
Dh&=&\left.\frac{\del h}{\del T}\right|_{\mu/T}DT+
\left.\frac{\del h}{\del(\mu/T)}\right|_TD(\frac{\mu}{T})
\ea
on the left hand sides of eqs.~(\ref{De}) and (\ref{Dh}) results in a coupled set
of equations for $DT$ and $D(\mu/T)$. These are easily solved to arrive at
\ba
DT&=&\frac{(P/n)\left[T-\left.\frac{\del h}{\del(\mu/T)}\right|_T\right]\del_\mu
u^\mu}{\left.\frac{\del e}{\del(\mu/T)}\right|_T
\left[hT^{-1}-\left.\frac{\del h}{\del T}\right|_{\mu/T}\right]-
\left.\frac{\del e}{\del T}\right|_{\mu/T}
\left[T-\left.\frac{\del h}{\del(\mu/T)}\right|_T\right]}\nn\\
D(\frac{\mu}{T})&=&\frac{-(P/n)\left[hT^{-1}-\left.\frac{\del h}{\del T}\right|_{\mu/T}\right]\del_\mu
u^\mu}{\left.\frac{\del e}{\del(\mu/T)}\right|_T
\left[hT^{-1}-\left.\frac{\del h}{\del T}\right|_{\mu/T}\right]-
\left.\frac{\del e}{\del T}\right|_{\mu/T}
\left[T-\left.\frac{\del h}{\del(\mu/T)}\right|_T\right]}~.
\label{DT}
\ea
We next evaluate the partial derivatives of $e$ and $h$ with respect $T$ and $(\mu/T)$
using the relations in Appendix-A. We get
\ba
\left.\frac{\del h}{\del T}\right|_{\mu/T}&=&z\left[5\frac{S_3^1}{S_2^1}+z\frac{S_2^0}{S_2^1}
-z\frac{S_3^1S_3^0}{(S_2^1)^2}\right]\nn\\
\left.\frac{\del e}{\del T}\right|_{\mu/T}&=&4z\frac{S_3^1}{S_2^1}+
z\frac{S_2^2S_3^0}{(S_2^1)^2}-\frac{S_2^2}{S_2^1}+z^2
\left[\frac{S_2^0}{S_2^1}-\frac{S_3^1S_3^0}{(S_2^1)^2}\right]\nn\\
\left.\frac{\del h}{\del(\mu/T)}\right|_T&=&Tz\left[\frac{S_3^0}{S_2^1}-
\frac{S_3^1S_2^0}{(S_2^1)^2}\right]\nn\\
\left.\frac{\del e}{\del(\mu/T)}\right|_T&=&-T\left[1-\frac{S_2^2S_2^0}{(S_2^1)^2}\right]
+Tz\left[\frac{S_3^0}{S_2^1}-\frac{S_3^1S_2^0}{(S_2^1)^2}\right]~.
\ea
Putting these in (\ref{DT}) we get
\ba
T^{-1}DT&=&(1-\gm')\del_\mu u^\mu\nn\\
TD(\frac{\mu}{T})&=&\left[(\gm''-1)h-\gm'''T\right]\del_\mu u^\mu
\ea
where
\be
\gamma'=\frac{(S_{2}^{0}/S_{2}^{1})^2-(S_{3}^{0}/S_{2}^{1})^2+4z^{-1}S_{2}^{0}S_{3}^{1}/(S_{2}^{1})^2+z^{-1}S_{3}^{0}/S_{2}^{1}}
{(S_{2}^{0}/S_{2}^{1})^2-(S_{3}^{0}/S_{2}^{1})^2+3z^{-1}S_{2}^{0}S_{3}^{1}/(S_{2}^{1})^2+2z^{-1}S_{3}^{0}/S_{2}^{1}-z^{-2}}
\ee
\be
\gamma''=1+\frac{z^{-2}}
{(S_{2}^{0}/S_{2}^{1})^2-(S_{3}^{0}/S_{2}^{1})^2+3z^{-1}S_{2}^{0}S_{3}^{1}/(S_{2}^{1})^2+2z^{-1}S_{3}^{0}/S_{2}^{1}-z^{-2}}
\ee
\be
\gamma'''=\frac{S_{2}^{0}/S_{2}^{1}+5z^{-1}S_{3}^{1}/S_{2}^{1}-S_{3}^{0}S_{3}^{1}/(S_{2}^{1})^2}
{(S_{2}^{0}/S_{2}^{1})^2-(S_{3}^{0}/S_{2}^{1})^2+3z^{-1}S_{2}^{0}S_{3}^{1}/(S_{2}^{1})^2+2z^{-1}S_{3}^{0}/S_{2}^{1}-z^{-2}}
\ee

We now come back to eq.~(\ref{tr_long}) in which we replace $DT$, $D(\frac{\mu}{T})$ and
$Du^\mu$ using eqs.~(\ref{DT}) and (\ref{Du}) to get
\ba
&&(p\cdot u) p_\mu\left[\frac{\nabla^\mu T}{T}-\frac{\nabla^\mu
P}{nh}\right]+Tp_\mu\nabla^\mu(\frac{\mu}{T})-p_\mu p_\nu\left[\nabla^\mu
u^\nu-\frac{1}{3}\De^\mn\nabla\cdot u\right]+\nn\\
&&\left[(p.u)^2(1-\gm')+p\cdot u\left\{(\gm''-1)h-\gm'''T\right\}-\frac{1}{3}p_\mu
p_\nu\De^\mn\right]\del\cdot u=-\frac{T\cl[\phi]}{f^{(0)}(1+f^{(0)})}
\ea
which on further simplification and rearrangement gives
\be
[Q\partial\cdot u+p_{\mu}\De^{\mu\nu}(p\cdot u- h)
(T^{-1}\partial_{\nu}T-Du_{\nu})-
\langle p_{\mu}p_{\nu} \rangle \langle \del^{\mu} u^{\nu}
\rangle]f^{(0)}(1+f^{(0)})
=-T\cl[\phi]
\ee
where the abbreviation
\be
Q=-\frac{1}{3}m^2+(p\cdot u)^2\{\frac{4}{3}-\gamma{'}
\}+p\cdot u\{(\gamma{''}-1) h
-\gamma{'''}T \}~.
\ee
and
\ba
\la\del^{\mu} u^{\nu}\ra&=&\left[\nabla^\mu
u^\nu-\frac{1}{3}\De^\mn\nabla\cdot u\right]\nn\\
\la p_\mu p_\nu\ra&=&p_\mu p_\nu-\frac{p^2}{3}\De_\mn-p\cdot u(p_\mu u_\nu+p_\nu
u_\mu-\frac{1}{3}g_\mn p\cdot u)+\frac{2}{3}(p\cdot u)^2u_\mu u_\nu~.
\ea

\section{Appendix C}
\setcounter{equation}{0}
\renewcommand{\theequation}{C.\arabic{equation}}

Here we briefly describe how the leading order expressions for the 
viscosities are obtained from the integral equations (\ref{AA}) and (\ref{CC}). 
Let us first consider the bulk viscosity $\zeta$. 
Following~\cite{Polak,Davesne} we multiply both sides of eq.~(\ref{AA})
by Laguerre polynomial of order 1/2 and degree $n$ and integrate
over $d\Gm_p$ to get
\be
[A(\tau),L_n^{{1}/{2}}(\tau)]=\frac{\alpha_n}{n}
\label{A_bra}
\ee
where $\tau=\beta(p\cdot u-m)$ and 
\be
\alpha_n=-\frac{1}{nT}\int d\Gm_p f^{(0)}(p)\{1+f^{(0)}(p)\}
QL_n^{{1}/{2}}(\tau)~.
\label{alpha_n}
\ee
and the abreviation
\be
[F,G]=\frac{1}{4n^2}\int d\Gm_p d\Gm_k d\Gm_{p'}\ d\Gm_{k'}f^{(0)}(p)f^{(0)}(k)
\{1+f^{(0)}(p')\}\{1+f^{(0)}(k')\}\de(F)\de(G)\ W
\label{bracket}
\ee
with
\ba
\de(F)&=&F(p)+F(k)+F(p')+F(k')\nn\\
\de(G)&=&G(p)+G(k)+G(p')+G(k')~.
\ea
Expanding $A$ as
\be
A(\tau)=\sum_{m=0}^{\infty}a_mL_m^{{1}/{2}} (\tau)
\ee
and putting in (\ref{A_bra}) the expression for bulk viscosity
to first order is obtained as
\be
\zeta=T\frac{\alpha_{2}^{2}}{a_{22}}
\ee
where
$\alpha_2$ is given by eq.~(\ref{alpha_n}) with $n=2$. 
The denominator $a_{22}$ in (\ref{zeta}) can be expressed using (\ref{bracket}) as
\be
a_{22}=[L_2^{{1}/{2}}(\tau),L_2^{{1}/{2}}(\tau)]
\label{a22}
\ee
and involves a 12-dimensional integral. This is considerably simplified by
making a change of variables as described in~\cite{Davesne,Polak}
reducing the number of integrations essentially to five so that
\ba
a_{22}&=&\frac{m^6}{2\pi^4n^2}\int_0^\infty d\chi \int_0^\infty d\psi \int_0^\pi d\theta
\int_0^{2\pi} d\phi\int_0^\pi  d\Theta\nn\\
&\times&f^{(0)}(p)f^{(0)}(k)\{1+f^{(0)}(p')\}\{1+f^{(0)}(k')\}\nn\\
&\times&\de\{L_2^{{1}/{2}}(\tau)\}\de\{L_2^{{1}/{2}}(\tau)\}
\frac{1}{2}\frac{d\sigma}{d\Omega}(\psi,\Theta)\nn\\
&\times&\sinh^2\chi\sinh^3\psi\cosh^3\psi\sin\Theta\sin\theta~.
\ea
Using now 
\[f^{(0)}(p)f^{(0)}(k)\{1+f^{(0)}(p')\}\{1+f^{(0)}(k')\}=
\frac{e^{-2\mu/T}e^{2z\cosh\psi\cosh\chi}}{(e^E-1)(e^F-1)(e^G-1)(e^H-1)}\]
where the exponents in the Bose functions
are given by
\ba
E&=&z(\cosh\psi\cosh\chi-\sinh\psi\sinh\chi\cos\theta)-\mu/T\nonumber\\
F&=&z(\cosh\psi\cosh\chi-\sinh\psi\sinh\chi\cos\theta')-\mu/T\nonumber\\
G&=&E+2z\sinh\psi\sinh\chi\cos\theta\nonumber\\
H&=&F+2z\sinh\psi\sinh\chi\cos\theta'
\ea
and
\be
\de\{L_2^{{1}/{2}}(\tau)\}\de\{L_2^{{1}/{2}}(\tau)\}=z^4(\sinh\psi\sinh\chi)^4
(\cos^2\theta-\cos^2\theta')^2
\ee
we finally get
\be
a_{22}=2z^2I_{3}(z)~.
\ee
The relative angle $\theta^\prime$ is defined by,
\(
\cos\theta'=\cos\theta\cos\Theta-\sin\theta\sin\Theta\cos\phi~.
\)
Also, on simplification, the integral $\alpha_2$ is finally given by
\ba
\alpha_{2}&=&\frac{z^3}{2} [\frac{1}{3}(\frac{S_{3}^0}{S_{2}^{1}}-z^{-1})
+(\frac{S_{2}^{0}}{S_{2}^{1}}+\frac{3}{z}\frac{S_{3}^{1}}{S_{2}^{1}})
\{(1-\gamma'')\frac{S_{3}^{1}}{S_{2}^{1}}+\gamma'''z^{-1}\}
\nonumber\\
&-&(\frac{4}{3}-\gamma')\{ \frac{S_{3}^{0}}
{S_{2}^{1}}+15z^{-2}\frac{S_{3}^{2}}{S_{2}^{1}}+2z^{-1}\}]~.
\ea

The calculation for the shear viscosity $\eta$ follows a similar procedure and is more involved
because of the tensorial nature of $C_\mn=C\la p_\mu p_\nu\ra$. In this case
eq.~(\ref{CC}) is multiplied by Laguerre polynomial of order $5/2$ on both sides
and then integrated over $d\Gm_p$ to get the formal expression
\be
[C(\tau)\la p_\mu p_\nu\ra,L_n^{{5}/{2}}(\tau)\la p^\mu
p^\nu\ra]=\frac{m}{n}T\gm_n
\label{C_bra}
\ee
where
\be
\gm_n=-\frac{1}{m nT^2}\int d\Gm_p f^{(0)}(p)\{1+f^{(0)}(p)\}
L_n^{{5}/{2}}(\tau)\la p_\mu p_\nu\ra\la p^\mu p^\nu\ra~.
\label{gaman}
\ee
Writing in this case 
\be
C(\tau)=\sum_{m=0}^{\infty}c_mL_m^{{5}/{2}} (\tau)
\ee
and putting in (\ref{C_bra}), the first approximation to $\eta$ is obtained as
\be
\eta=\frac{T}{10}\ \frac{\gamma_0^2}{c_{00}}
\ee
where
\be
c_{00}=\frac{1}{m^2 T^2}[L_0^{{5}/{2}}(\tau)\la p_\mu p_\nu\ra,
L_0^{{5}/{2}}(\tau)\la p^\mu p^\nu\ra]~.
\ee
Simplification along the lines of~\cite{Davesne,Polak} finally yields
\ba
\gamma_0&=&-10\frac{S_3^{2}(z)}{S_2^{1}(z)}~,\nonumber\\
c_{00}&=&16I_1(z)+16I_2(z)+\frac{16}{3}I_3(z)~.
\ea
The integrals $I_\alpha(z)$ are defined as
\ba
I_\alpha(z)&=&\frac{z^4}{[S_2^{1}(z)]^2} \ e^{(-2\mu/T)}\int_0^\infty d\psi\ \cosh^3\psi
\sinh^7\psi\int_0^\pi
d\Theta\sin\Theta\frac{1}{2}\frac{d\sigma}{d\Omega}(\psi,\Theta)\int_0^{2\pi}
d\phi\nonumber\\&&\int_0^\infty d\chi \sinh^{2\alpha} \chi
\int_0^\pi d\theta\sin\theta\frac{e^{2z\cosh\psi\cosh\chi}}
{(e^E-1)(e^F-1)(e^G-1)(e^H-1)}\ M_\alpha(\theta,\Theta)
\ea
where the functions $M_\alpha$ stand for
\ba
M_1(\theta,\Theta)&=&1-\cos^2\Theta~,\nonumber\\M_2(\theta,\Theta)&=&\cos^2\theta+\cos^2\theta'
-2\cos\theta\cos\theta'\cos\Theta~,\nonumber\\
M_3(\theta,\Theta)&=&[\cos^2\theta-\cos^2\theta']^2
\ea


\begin{thebibliography}{99}

\bibitem{KSS}
P.~Kovtun, D.~T.~Son and A.~O.~Starinets,
Phys.\ Rev.\ Lett.\  {\bf 94}, 111601 (2005).  

\bibitem{Csernai}
L.~P.~Csernai, J.~I.~Kapusta and L.~D.~McLerran,
Phys.\ Rev.\ Lett.\  {\bf 97}, 152303 (2006). 

\bibitem{Rebhan}
A.~Rebhan and D.~Steineder,
Phys.\ Rev.\ Lett.\  {\bf 108} (2012) 021601

\bibitem{Cheng}
M.~Cheng {\it et al.},
Phys.\ Rev.\ D {\bf 81} (2010) 054504

\bibitem{Kharzeev}
D.~Kharzeev and K.~Tuchin,
JHEP {\bf 0809} (2008) 093

\bibitem{Dobado1}
A.~Dobado, F.~J.~Llanes-Estrada and J.~M.~Torres-Rincon,
Phys.\ Rev.\ D {\bf 80} (2009) 114015

\bibitem{Dobado2}
A.~Dobado and J.~M.~Torres-Rincon,
Phys.\ Rev.\ D {\bf 86} (2012) 074021

\bibitem{Chen2}
J.~-W.~Chen and J.~Wang,
Phys.\ Rev.\ C {\bf 79} (2009) 044913

\bibitem{Zubarev}
D. ~N.~Zubarev, {\it Non-equilibrium Statistical Thermodynamics} (Consultants
Bureau, NN, 1974).

\bibitem{Sakagami}
A.~Hosoya, M.~a.~Sakagami and M.~Takao,
Annals Phys.\  {\bf 154} (1984) 229.

\bibitem{Lang}
R.~Lang, N.~Kaiser and W.~Weise,
Eur.\ Phys.\ J.\ A {\bf 48} (2012) 109

\bibitem{Mallik_kubo}
S.~Mallik and S.~Sarkar,
arXiv:1211.2588 [nucl-th].

\bibitem{Jeon}
S.~Jeon and L.~G.~Yaffe,
Phys.\ Rev.\  D {\bf 53}, 5799 (1996).  

\bibitem{Weinberg1}
S.~Weinberg,
Physica A {\bf 96} (1979) 327.  

\bibitem{Gasser}
J.~Gasser and H.~Leutwyler,
Annals Phys.\  {\bf 158}, 142 (1984). 

\bibitem{Santalla}
A.~Dobado and S.~N.~Santalla,
Phys.\ Rev.\  D {\bf 65}, 096011 (2002).

\bibitem{Chen}
J.~W.~Chen, Y.~H.~Li, Y.~F.~Liu and E.~Nakano,
Phys.\ Rev.\  D {\bf 76}, 114011 (2007)  

\bibitem{Dobado3}
A.~Dobado and F.~J.~Llanes-Estrada,
Phys.\ Rev.\  D {\bf 69}, 116004 (2004).  

\bibitem{Itakura}
K.~Itakura, O.~Morimatsu and H.~Otomo,
Phys.\ Rev.\  D {\bf 77}, 014014 (2008).   

\bibitem{Prakash}
M.~Prakash, M.~Prakash, R.~Venugopalan and G.~Welke,
Phys.\ Rept.\  {\bf 227} (1993) 321.

\bibitem{Davesne}
D.~Davesne,
Phys.\ Rev.\  C {\bf 53}, 3069 (1996).

\bibitem{Moore}
E.~Lu and G.~D.~Moore,
Phys.\ Rev.\ C {\bf 83} (2011) 044901

\bibitem{Fraile}
D.~Fernandez-Fraile and A.~Gomez Nicola,
Phys.\ Rev.\ Lett.\  {\bf 102} (2009) 121601

\bibitem{Bertsch1}
G.~Bertsch, M.~Gong, L.~D.~McLerran, P.~V.~Ruuskanen and E.~Sarkkinen,
Phys.\ Rev.\  D {\bf 37} (1988) 1202.	

  
\bibitem{Dobado_kss}
A.~Dobado and F.~J.~Llanes-Estrada,
Eur.\ Phys.\ J.\ C {\bf 49} (2007) 1011


\bibitem{Sukanya}
S.~Mitra, S.~Ghosh and S.~Sarkar,
Phys.\ Rev.\ C {\bf 85} (2012) 064917

\bibitem{Purnendu}
P.~Chakraborty and J.~I.~Kapusta,
Phys.\ Rev.\  C {\bf 83} (2011) 014906.

\bibitem{Polak}
W.~A.~Van Leeuwen, P.~H.~Polak and S.~R.~De Groot,
Physica {\bf 66}, 455 (1973).   

\bibitem{Welke}
G.~M.~Welke, R.~Venugopalan and M.~Prakash,
Phys.\ Lett.\ B {\bf 245} (1990) 137.

\bibitem{Anantha}
B.~Ananthanarayan, G.~Colangelo, J.~Gasser and H.~Leutwyler,
Phys.\ Rept.\  {\bf 353} (2001) 207

\bibitem{Weinberg_PRL}
S.~Weinberg,
Phys.\ Rev.\ Lett.\  {\bf 17} (1966) 616.



\bibitem{Bellac} M. Le Bellac, {\it Thermal Field Theory} (Cambridge University
Press, Cambridge, 1996).  

\bibitem{Mallik_RT}  S.~Mallik and S.~Sarkar,
Eur.\ Phys.\ J.\  C {\bf 61}, 489 (2009).

\bibitem{Ghosh1}  S.~Ghosh, S.~Sarkar and S.~Mallik,
Eur.\ Phys.\ J.\  C {\bf 70}, 251 (2010).

\bibitem{Weldon} H.~A.~Weldon,
Phys.\ Rev.\  D {\bf 28} (1983) 2007.

\bibitem{PDG}
K.~Nakamura {\it et al.}  [Particle Data Group],
J.\ Phys.\ G {\bf 37} (2010) 075021.

\bibitem{Ghosh2}
S.~Ghosh and S.~Sarkar,
Nucl.\ Phys.\  A {\bf 870-871}, 94 (2011)  
	
\bibitem{Bertsch2}  
H.~W.~Barz, H.~Schulz, G.~Bertsch and P.~Danielewicz,
Phys.\ Lett.\  B {\bf 275} (1992) 19.

\bibitem{Hirano}
T.~Hirano and K.~Tsuda,
Phys.\ Rev.\ C {\bf 66} (2002) 054905

\bibitem{Bebie}
H.~Bebie, P.~Gerber, J.~L.~Goity and H.~Leutwyler,
Nucl.\ Phys.\ B {\bf 378} (1992) 95.

\bibitem{Nicola}
D.~Fernandez-Fraile and A.~Gomez Nicola,
Phys.\ Rev.\ D {\bf 80} (2009) 056003

\bibitem{Gerber}
P.~Gerber and H.~Leutwyler,
Nucl.\ Phys.\ B {\bf 321} (1989) 387.

\bibitem{Dobado_bulk}
A.~Dobado, F.~J.~Llanes-Estrada and J.~M.~Torres-Rincon,
Phys.\ Lett.\ B {\bf 702} (2011) 43

\bibitem{Nakano}
E.~Nakano,
hep-ph/0612255.

\bibitem{Dusling}
K.~Dusling and T.~Schafer,
Phys.\ Rev.\ C {\bf 85} (2012) 044909


\end{thebibliography}
\end{document}